\documentclass[preprint,showpacs,preprintnumbers,amsmath,amssymb,floatfix]{revtex4}
\usepackage{graphicx}

\begin{document}


\title{Temperature-dependent contact resistances in high-quality polymer 
field-effect transistors}

\author{B.H. Hamadani and D. Natelson}

\affiliation{Department of Physics and Astronomy, Rice University, 6100 Main St., Houston, TX 77005}

\date{\today}

\begin{abstract}
Contact resistances between organic semiconductors and metals can
dominate the transport properties of electronic devices incorporating
such materials.  We report measurements of the parasitic contact
resistance and the true channel resistance in bottom contact
poly(3-hexylthiophene) (P3HT) field-effect transistors with channel
lengths from 400~nm up to 40~$\mu$m, from room temperature down to
77~K.  For fixed gate voltage, the ratio of contact to channel
resistance decreases with decreasing temperature.  We compare this
result with a recent model for metal-organic semiconductor contacts.
Mobilities corrected for this contact resistance can approach
1~cm$^{2}$/Vs at room temperature and high gate voltages.
\end{abstract}

\maketitle

Much progress has been made in recent years in the development of
field-effect transistors based on organic semiconductors (OFETs)
\cite{DimitrakopoulosetAl01IBM}.  Such devices have attracted interest
for their potential applications in inexpensive and flexible
electronics.  Improving the device characteristics of such OFETs to
have higher ``on'' conductances and larger on-off ratios in
solution-processable materials is a technological priority.

The parasitic series resistance, $R_{\rm s}$, between the organic
semiconductor (OSC) and the metal electrodes is of particular
interest\cite{Scott03JVSTA}.  Contacts in OFETs (extremely restricted
contact geometry, essentially undoped OSCs) are distinct from those
in, for example, organic light emitting diodes (large contact areas,
highly doped OSCs).  Contact resistance is often neglected when
inferring the mobility from transistor characteristics.
Different approaches have been developed to differentiate between
contact and channel resistances, including analyses of single device
characteristics\cite{HorowitzetAl99JAP,StreetetAl02APL}, scanning
potentiometry\cite{SeshadrietAl01APL,BurgietAl02APL}, and scaling of
total resistance with channel length in a series of
devices\cite{KaluketAl03SSE,NecludiovetAl03SSE,ZaumsiletAl03JAP,MeijeretAl03APL}.
Such experiments have already shown that $R_{\rm s}$ can easily
dominate $R_{\rm ch}$ in micron-scale pentacene
OFETs\cite{KaluketAl03SSE,NecludiovetAl03SSE}, and that $R_{\rm s}$
correlates inversely with mobility in polymer OFETs at room
temperature\cite{MeijeretAl03APL}.

The physics relevant to $R_{\rm s}$ at metal-OSC interfaces is a
subject of much discussion\cite{Scott03JVSTA}.  Models include
Schottky contacts\cite{StreetetAl02APL} and antiparallel Schottky
diodes in parallel with a
resistance\cite{KaluketAl03SSE,NecludiovetAl03SSE}.  More
sophisticated treatments include the fact that conduction in
disordered OSCs is by hopping\cite{AbkowitzetAl98JAP}, image charge
effects\cite{ArkhipovetAl98JAP}, and account for charge recombination
at the metal-OSC interface\cite{ScottetAl99CPL}.  This last model has
received recent experimental support in studies that examine scaling
of the contact resistivity with
mobility\cite{ShenetAl01PRL,MeijeretAl03APL}.

We report transport measurements in set of OFETs made from solution
cast, regio-regular poly(3-hexylthiophene) (P3HT), with channel
lengths from 400~nm to 40~$\mu$m.  We determine the parasitic series
contact resistance, $R_{\rm s}$, and the true channel resistance,
$R_{\rm ch}$, from the dependence of the total source-drain
resistance, $R_{\rm on}$, on the channel length, $L$.  Both $R_{\rm
s}$ and $R_{\rm ch}$ increase as temperature is decreased, with
$R_{\rm ch}$ varying more rapidly.  The result is that the ratio
$R_{\rm s}/R_{\rm ch}$ actually {\it decreases} as $T$ is lowered;
relative to the mobility, the contacts actually improve as $T$ is
decreased.  We compare this trend with a theoretical treatment of
metal-OSC contacts, and within that model are able to estimate the
Schottky barrier for hole injection from Au into P3HT.  We also note
that our highest quality samples have field-effect mobilities,
$\mu_{\rm FE}$, (corrected for contact effects) approaching
1~cm$^{2}$/Vs, with typical values $\sim 0.2$ cm$^{2}$/Vs.  

Devices are made in a bottom-contact configuration (see
Fig.~\ref{fig:mobT}a inset) on a degenerately doped $p+$ silicon wafer
used as a gate.  The gate dielectric is 200~nm of thermal SiO$_{2}$.
Source and drain electrodes are patterned using electron beam
lithography.  The electrodes are deposited by electron beam
evaporation of 4~nm Ti and 25~nm of Au followed by liftoff.  The
substrates are then cleaned for one minute in a 1:1 solution of
NH$_{4}$OH:H$_{2}$O$_{2}$ (30\%), rinsed in deionized water, and
exposed for one minute to an oxygen plasma.

The organic semiconductor is 98\% regio-regular P3HT\cite{purchase}, a
well studied
material\cite{BaoetAl96APL,SirringhausetAl98Science,SirringhausetAl99Nature}.
As-received RR-P3HT is dissolved in chloroform at a 0.02\% weight
concentration, passed through PTFA 0.2~$\mu$m filters, and solution
cast\cite{BaoetAl96APL} onto the cleaned substrates, with the solvent
allowed to evaporate in ambient conditions.  The resulting film
thicknesses are tens of nm as determined by atomic force microscopy
(AFM). Casting produces nonuniform films, though thickness variations
do not affect device performance detectably.  Variations in field
effect mobilities may result from subtle differences in deposition
condutions.  All devices are stored in vacuum dessicators until use.
The measurements are performed in vacuum ($\sim 10^{-6}$~Torr) in a
variable-temperature probe station using a semiconductor parameter
analyzer (HP4145B).  Prior to measurement, the samples are heated in
the probe station to 330~K under vacuum for up to 24 hours, to remove
any residual solvent and unintended doping due to exposure to
atmosphere.  On-off ratios at 300~K in the linear regime are typically
several hundred.  We report data for three arrays, each with at least
eight FET devices, with parameters as described in
Table~\ref{tab:samples}.  For brevity, the data in subsequent figures
are drawn from the $w = 5~\mu$m ensemble, and are representative of
the other channel widths.

The devices operate as standard $p$-type FETs in accumulation
mode\cite{BaoetAl96APL,SirringhausetAl98Science,SirringhausetAl99Nature,AleshinetAl01SM,MeijeretAl02APL}.
With the source electrode as ground, the devices are measured in the
shallow channel regime ($V_{D}<V_{G}$).  For each device at each gate
voltage, the linear portion of $I_{D}-V_{D}$ is used to find $R_{\rm
on}\equiv \partial I_{D}/\partial V_{D}$, the total source-drain
resistance.  The values of $R_{\rm on}$ are then plotted as a function
of channel length for the ensemble of OFETs, as shown in
Fig.~\ref{fig:RvsL}.  The slope of such a plot describes $R_{\rm ch}$
per unit channel length.  The intercept (the extrapolated resistance
of a device of zero channel length) gives $R_{\rm s}$, the total
parasitic series resistance of the source and drain contacts.  The
true field-effect mobility, $\mu_{\rm FE}$, may be inferred from the
gate voltage dependence of $R_{\rm on}$:
\begin{equation}
\frac{\partial\left[ \left(\frac{\partial R_{\rm on}}{\partial L}\right)^{-1}\right]}{\partial V_{G}} = \mu_{\rm FE}(V_{G},T)wC_{\rm ox},
\label{eq:mobility}
\end{equation}
where $C_{\rm ox}$ is the capacitance per unit area of the gate oxide.
Mobilities inferred from the saturation regime ({\it uncorrected} for
contact effects) are systematically lower than corrected $\mu_{\rm
FE}$ values, and are comparatively independent of $L$.  Device $IV$
characteristics are stable with thermal cycling, and samples stored in
vacuum for two months exhibit undegraded performance.


Figure~\ref{fig:mobT}a shows the temperature dependence of $\mu_{\rm
FE}$ for this series of 5~$\mu$m wide devices as a
function of $T^{-1}$.  The temperature dependence is well approximated
as thermal activation, with the activation energy, $\Delta$, weakly
dependent on gate voltage.  For $V_{G}=-90$~V, $\Delta \approx
29.4$~meV; for $V_{G}=-30$~V, $\Delta \approx 50.8$~meV.

The temperature dependence of the parasitic contact resistance for the
same devices is shown in Figure~\ref{fig:mobT}b.  The contact
resistance increases significantly as $T$ is decreased, again in an
activated fashion.  The activation energies are very similar to those
for $\mu_{\rm FE}$, strongly suggesting that the same physics couples
both these parameters.  The activation energies inferred for $R_{\rm
s}$ are systematically {\it below} those inferred for $\mu_{\rm FE}$
for all gate voltages by a few meV ($\sim$~4~meV for the 5~$\mu$m
ensemble of devices).

A recently developed theory of OSC-metal
contacts\cite{ScottetAl99CPL,ShenetAl01PRL} based on earlier work
examining injection into poor conductors\cite{EmtageetAl66PRL}
suggests why this should be so.  Scott and
Malliaras\cite{ScottetAl99CPL} predict that the rate of injection into
an OSC from a contact limited metal electrode is proportional to the
OSC mobility.  In particular they show
\begin{equation}
J_{\rm INJ}= 4 \psi^{2} N_{0} e \mu E \exp(-\phi_{B}/k_{\rm B}T)\exp(f^{1/2}),
\label{eq:scott1}
\end{equation} 
where $\psi$ is a slowly varying function of electric field, $E$;
$N_{0}$ is the density of localized sites available for hopping
conduction; $\phi_{B}$ is the Schottky barrier between the metal and
the OSC; and the $f = e^3 E/[4 \pi \epsilon \epsilon_{0}(k_{\rm
B}T)^{2}]$ term is due to Schottky barrier lowering.  This variation
of injection current density with OSC mobility has been confirmed in
two-terminal OSC-metal diodes\cite{ShenetAl01PRL}.  In the low field
limit, if the mobility itself is thermally activated with a
characteristic energy $\Delta(V_{G})$, one would expect
$R_{s} \propto \exp{(\Delta(V_{G}) + \phi_{B})/k_{\rm B}T}$.
The similarity of activation energies for $R_{\rm s}$ and $\mu_{\rm FE}$
would follow naturally, provided that the Schottky barrier between
Au and P3HT is low.

Note that Eq.~(\ref{eq:scott1}) implies that, for a given system at a
fixed temperature, $R_{\rm s} \propto 1/\mu_{\rm FE}$.  The constant
of proportionality is temperature-dependent, and would be dominated by
the Schottky barrier contribution, $\sim \exp(\Delta/k_{\rm B}T)$.
Figure~\ref{fig:Vgfig} is a plot of $R_{\rm s}$ vs. $\mu_{\rm FE}$ for
{\it all three device ensembles}, including data for {\it all gate
voltages and temperatures} examined.  The fit demonstrates that
$R_{\rm s} \sim 1/\mu_{\rm FE}^{1.09}$ over {\it four decades} of
mobility.  This strongly supports the mobility dependence of the
injection model of Eq.~(\ref{eq:scott1}) derived in
Ref.~\cite{ScottetAl99CPL}, provided the Schottky barrier for the
Au/P3HT interface is nearly zero.  Such a small barrier is consistent
with the similarity in activation energies for $R_{\rm s}$ and
$\mu_{\rm FE}$ described above and seen in Fig.~\ref{fig:mobT}.

Fig.~\ref{fig:ratiovsT} shows the ratio of $R_{\rm s}$ to $R_{ch}$ for
a $L = 1~\mu$m device from the 5~$\mu$m wide ensemble of devices as a
function of temperature.  Error bars are significant because of the
uncertainty in the slope and intercept parameters obtained from data
like that in Fig.~\ref{fig:RvsL}.  These errors are dominated by
device-to-device fluctuations within the ensemble.  The ratio {\it
decreases} slowly as $T$ is reduced.  Within the model of
Eqs.~(\ref{eq:scott1}), this suggests that the barrier height for our
Au-P3HT interface is actually slightly {\it negative}, again
consistent with the systematic difference in activation energies
discussed above.  Relative to the channel, the contacts actually {\it
improve} slightly at low temperatures, so that a device that is bulk
limited at room temperature will remain so at lower temperatures.


We have used the length dependence of the channel resistance to
extract the intrinsic mobility and parasitic contact resistance as a
function of temperature and gate voltage for several series of
bottom-contact, solution-cast polymer OFETs.  The temperature and gate
voltage dependence of the contact resistance and the inferred mobility
support a recently developed model of charge injection from metals
into disordered OSCs.  We find that the ratio of contact to channel
resistance actually {\it decreases} slightly as $T$ is reduced,
implying a very small negative Schottky barrier for the gold-P3HT
interface in these devices.  Once parasitic contact resistances are
taken into account, the mobility of solution-cast P3HT can
approach 1~cm$^{2}$/Vs at room temperature, nearly an order of
magnitude larger than uncorrected mobilities.  These results indicate
that performance of P3HT-based OFETs can be limited more by contact
physics than by the intrinsic tranport physics in the polymer itself.

The authors gratefully acknowledge the support of the Robert A. Welch
Foundation and the Research Corporation.

\pagebreak

\begin{figure}[h!]
\caption{\small $R_{\rm on}$ as a function of $L$ at 300~K for a series of P3HT OFETs with
channel widths of 5~$\mu$m.   
Inset:  $I_{D}$ vs. $V_{D}$ for several gate voltages (top to bottom, -90~V, -70~V, -50~V, -30~V, -10~V) in the $L = 2~\mu$m device from this series at $T = 210~$K.  From -90~V to -10~V, the on-off ratio for this device is 4.1$\times 10^3$.}
\label{fig:RvsL}
\end{figure}

\begin{figure}[h!]
\caption{\small (a)  Mobility as a function of $1/T$ for several 
gate voltages, found via Eq.~(\ref{eq:mobility}) in a series of 5~$\mu$m wide devices.  Note the high values of $\mu_{\rm FE}$ at large $V_{G}$ and 
high $T$.
(b) Parasitic contact resistance as a function of $1/T$ for the same
devices.}
\label{fig:mobT}
\end{figure}

\begin{figure}[h!]
\caption{\small 
A summary plot of contact resistivity as a function of
field-effect mobility, for all three ensembles of devices, and for all
gate voltages and temperatures examined.  Error bars have been omitted
for clarity.  The fit is to a power law with exponent -1.09.  
Inset:$R_{\rm s}(V_{G})$ vs. $R_{\rm
ch}(V_{G})\times V_{G} (\propto 1/\mu_{\rm FE})$ at 210~K (upper)
and 100~K (lower) for the 5~$\mu$m channel width ensemble of devices.  
The linear dependences
confirm that $R_{\rm s} \sim \mu_{\rm FE}^{-1}$, as predicted
in the model of Ref.~{\protect{\cite{ScottetAl99CPL}}}. }
\label{fig:Vgfig}
\end{figure}

\begin{figure}[h!]
\caption{\small 
$R_{\rm s}/R_{\rm ch}$ as a function of temperature for several gate
voltages in the $w = 5~\mu$m devices, for a channel length of
1~$\mu$m.  Relative to the channel, the contacts actually {\it
improve} as the temperature decreases.}
\label{fig:ratiovsT}
\end{figure}

\clearpage

\begin{table}
\caption{\small Parameters describing ensembles of devices analyzed in
this study.  Mobility values are obtained from $I_{D}-V_{D}$ data
using Eq.~(\ref{eq:mobility}).  Contact resistivity values are
computed by extrapolating $R_{\rm on}$ back to $L = 0$. 
*Data for the 30 micron wide devices were at $V_{G}=-70$~V
rather than -90~V.}
\vspace{4mm}
\begin{tabular}{|c c c c c|}
\hline
~~~~Ensemble~~~~  & ~~~~min. $L~[\mu$m]~~~~ & ~~~~max. $L~[\mu$]~~~~ & ~~~~$\mu_{\rm FE}$ (300~K,~~~~  & ~~~~$R_{\rm s}w$ (300~K,~~~~ \\
width [$\mu$m] & & & $V_{G}=-90$~V) & $V_{G}=-90$~V)\\
 & & & [cm$^{2}$/Vs] & [$\Omega$-cm] \\
\hline
5 & 0.4 & 5 & 0.73 & $1.1 \times 10^3$ \\
30* & 2 & 30 & 0.16 & $9.6 \times 10^3$ \\
100 & 5 & 40 & 0.13 & $1.1 \times 10^4$ \\
\hline
\end{tabular}
\label{tab:samples}
\vspace{-3mm}
\end{table}

\clearpage

\setcounter{figure}{0}

\begin{figure}
\begin{center}
\includegraphics[width=7cm, clip]{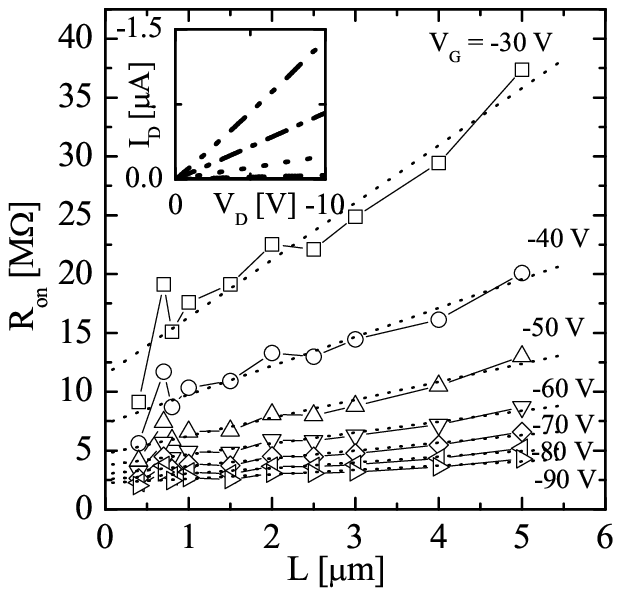}
\end{center}
\vspace{1.5cm}
\caption{of 4, B.H. Hamadani and D. Natelson ``Temperature dependent contact resistances in high quality polymer field effect transistors'', to appear in {\it Appl. Phys. Lett.}} 
\end{figure} 
  
\clearpage

\begin{figure}
\begin{center}
\includegraphics[width=7cm, clip]{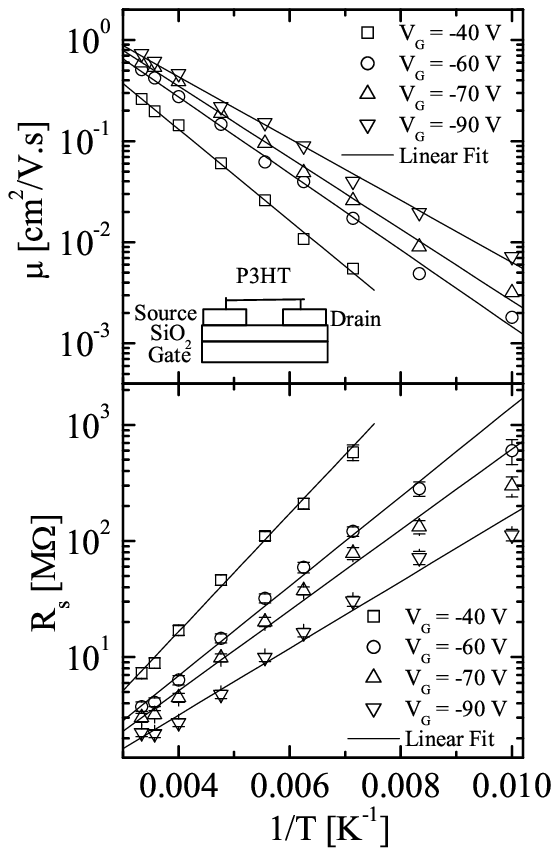}
\end{center}
\vspace{1.5cm}
\caption{of 4, B.H. Hamadani and D. Natelson ``Temperature dependent contact resistances in high quality polymer field effect transistors'', to appear in {\it Appl. Phys. Lett.}} 
\end{figure} 

\clearpage

\begin{figure}
\begin{center}
\includegraphics[width=7cm, clip]{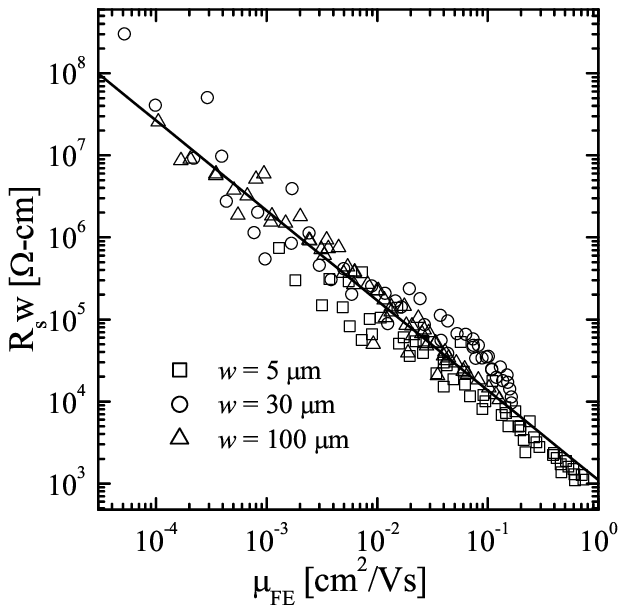}
\end{center}
\vspace{1.5cm}
\caption{of 4, B.H. Hamadani and D. Natelson ``Temperature dependent contact resistances in high quality polymer field effect transistors'', to appear in {\it Appl. Phys. Lett.}} 
\end{figure} 

\clearpage

\begin{figure}
\begin{center}
\includegraphics[width=7cm, clip]{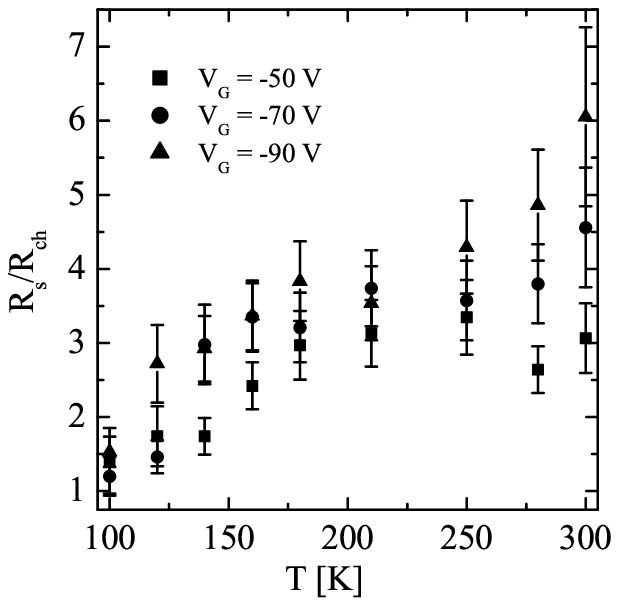}
\end{center}
\vspace{1.5cm}
\caption{of 4, B.H. Hamadani and D. Natelson ``Temperature dependent contact resistances in high quality polymer field effect transistors'', to appear in {\it Appl. Phys. Lett.}} 
\end{figure}

\end{document}